# A Review of How Supermassive Black Hole Binaries Can Be Detected and Characterized through Gravitational Lensing of Background Stars

Mu'allim Yakubu

Department of Physics, Air Force Institute of Technology Kaduna, Nigeria

Corresponding author: Yakubumuallim@gmail.com

***Abstract***

*Supermassive black hole binaries are an inevitable outcome of hierarchical galaxy mergers, but their sub-parsec evolutionary phase between the kpc separations resolvable by direct imaging and the millihertz gravitational-wave regime targeted by LISA remains observationally elusive. Gravitational lensing of background stars by SMBH binaries has emerged as a powerful complement to periodic variability searches, periodic Doppler-boost signals, and circumbinary accretion hydrodynamic models. Two distinct lensing channels are reviewed: (i) the gravitational lensing of Milky Way bulge stars by the central Sgr A* binary, whose secondary image is now within reach of ELT-class instruments; and (ii) the recently introduced Quasi-Periodic Lensing of Starlight mechanism, in which the rotation of caustics produced by a non-active SMBH binary magnifies individual bright stars in its host galaxy on the orbital period of the binary. The QPLS framework predicts 1-50 (190-5,000) $n_\star/pc^{-3}$ binaries with periods below 10 yr and $z < 0.3$, opening a multimessenger window onto pulsar-timing-array sources and LISA-band events. After outlining the theoretical foundation, microlensing formalism, observational status and the synergy with pulsar-timing arrays and LISA, we identify outstanding challenges and project the field into the LSST era.*

## 1. Introduction

Supermassive black holes are central-monolith components of massive galaxies and grow primarily through successive mergers (Burke-Spolaor et al., 2019; Merritt & Milosavljević, 2005). Each merger delivers two SMBHs to a common nucleus, where they are expected to harden through dynamical friction, three-body scattering with stars, and finally gravitational-wave emission (Holley-Bockelmann & Khan, 2015; Merritt & Milosavljević, 2005). The phase between ≈1 pc and ≈0.01 pc the notorious "final parsec" is where observational confirmation has lagged most acutely, since the binary is too compact for direct imaging yet radiates too weakly as a GW source for LISA-band resolution (Auclair et al., 2023; Bogdanović et al., 2022). Hierarchical structure-formation models imply the existence of large numbers of sub-parsec binaries at any instant, but the population synthesis predictions carry order-of-magnitude uncertainties that currently cannot be tested empirically (Burke-Spolaor et al., 2019; Mingarelli et al., 2017).

Electromagnetic signatures of SMBHBs have proliferated over the past decade. Periodic optical variability in large quasar surveys has yielded ≈111 candidates in the Catalina Real-time Transient Survey (Graham et al., 2015) and additional samples in PTF, Pan-STARRS, CRTS, ZTF and Gaia (Burke-Spolaor et al., 2019; Charisi et al., 2016; Kelley et al., 2021). The 12.2-yr twin-flare light curve of the BL Lac object OJ 287 has been modelled as repeated impacts of the secondary on the primary accretion disk, with mass and timing predictions confirmed by the 2015 and 2019 superflares (MacFadyen & Milosavljević, 2008; Valtonen et al., 2019). The periodicity of PG 1302-102 was initially attributed to relativistic Doppler boosting (D'Orazio et al., 2015) and to circumbinary accretion variability modulated on the orbital period (Liao et al., 2020). None of these methods, however, exploits the geometry of the binary itself to provide a direct probe of its orbit. Gravitational lensing by SMBHs offers a fundamentally geometric approach. Paczyński laid the foundation in 1996 by formalising single-lens and binary-lens microlensing in the local Universe (Paczyński, 1996), and Alexander subsequently showed that the lensing cross-section of an individual star near a massive black hole is dominated by the black hole's external shear and can develop the rich caustic topology of a binary deflector (Alexander, 2005). Building on this, Michałowski & Mróz demonstrated that stars in the Milky Way bulge can be lensed by Sgr A* into secondary images bright enough for detection by the ELT-class instruments of the next decade (Michałowski & Mróz, 2021). Most recently, Wang, Zumalacárregui & Kocsis introduced the Quasi-Periodic Lensing of Starlight mechanism, an extension of binary-lens theory in which the orbital rotation of the binary caustic amplifies background stars quasi-periodically on the orbital period of the SMBH binary itself (Hanxi et al., 2025; Wang et al., 2025). Because the binary caustic is rotationally swept past a dense stellar caustic pattern on the orbital timescale, QPLS light curves encode the binary mass, separation and eccentricity at a signal-to-noise set by the local stellar density $n_\star$ (Wang et al., 2025).

This review synthesises the theoretical and observational status of the technique, situates it alongside existing methods (periodic variability, Doppler boosting, circumbinary accretion, hydrodynamics), and identifies the next experimental steps.

## 2. Methodology

### 2.1 Microlensing framework

Microlensing by a single point mass of mass M produces an Einstein radius $\theta_E = \sqrt{\frac{4GM\ Ds\ Dls}{c^2\ Dl,}}$, with $D_l$, $D_s$ and $D_{ls}$ the angular-diameter distances to the lens, the source and the lens-to-source, respectively (Paczyński, 1996). The symmetric two-image geometry of the single-lens case is replaced in the binary-lens case by a caustic network associated with the inner Lagrangian points and a critical curve that breaks the degeneracy of the single-lens problem (Paczyński, 1996). The amplification diverges at the caustics and is of order unity in their vicinity.

For a star passing near a massive black hole ($M_\bullet$) of mass $M_\star \lesssim 10^{-6}\ M_\bullet$ in projection, the lensing cross-section of the star is $\theta^2_E\ (M_\star)/\theta^2_E\ (M_\bullet) \approx M_\star/M_\bullet \approx 10^{-6}$ times smaller than that of the black hole (Alexander, 2005). However, the shear from the black hole distorts the stellar caustic into a complex, radially elongated topology whose area is enhanced by up to an order of magnitude. As a star orbits the black hole, this elongated caustic scans the lens plane, and the chance of intersecting one of the two geodetic images of a background source becomes non-negligible (Alexander, 2005). The image splits into 2 or 4 sub-images with sub-Einstein-ring separations that are individually unresolvable but whose combined flux is significantly magnified (Alexander, 2005). Michałowski & Mróz used this framework to compute the limiting K-band magnitudes required to detect the secondary image of a bulge star lensed by Sgr A*, finding that resonant separations allow detections down to the magnitude of 21-23 mag for an isolated SMBH and even fainter for a binary SMBH (Michałowski & Mróz, 2021).

### 2.2 The QPLS mechanism

In QPLS, an SMBH binary is the lens and the sources are individual bright stars in the host galaxy that lie behind it. The binary deflector produces a network of rotating caustics whose inner-curve morphology depends on the binary mass ratio, separation and orbital phase (Hanxi et al., 2025; Wang et al., 2025). As the binary orbits, the caustic pattern sweeps across the observation plane. Source stars at small impact parameters are transiently magnified to extreme amplification when they encounter the rotating caustic, producing a light curve whose dominant peak-to-peak timescale is set by the binary orbital period (Hanxi et al., 2025; Wang et al., 2025). The mechanism is most efficient when the binary is equatorial-on (nearly edge-on) with respect to the line of sight, the stellar density of the host galaxy is large, and the binary separation is small enough that caustic crossings occur within typical survey cadences (Wang et al., 2025).

The morphology of the QPLS light curve differs depending on the dynamical regime. For tight, GW-dominated binaries (period $<10$ yr; total periods $<40$ yr with (Wang et al., 2025)), the stellar caustic crossings are rapid, large-amplitude magnification events that recur quasi-periodically on the orbital period (Wang et al., 2025). For wider, stellar-driven binaries (period $\gtrsim 10^4$ yr), caustic crossings driven by the orbital motion of the ambient stars occur as single or double flares superimposed on a long-timescale baseline (Wang et al., 2025). In all cases, the lensing magnification is of order tens to hundreds, far exceeding the $\sim$10% Doppler-boost amplitude typical of unequal-mass binaries, making it a more robust tracer against the 10%-level damped-random-walk variability of AGN accretion disks (Burke-Spolaor et al., 2019; Kelley et al., 2021).

### 2.3 Survey observables

The observational target of the methodology is the host galaxy light curve time series. Existing and upcoming photometric surveys Pan-STARRS, PTF, CRTS, ZTF, Gaia, Kepler, and the Vera C. Rubin Observatory's Legacy Survey of Space and Time provide global photometric coverage of tens of millions of galaxies with cadences ranging from days to weeks (Anonymous et al., 2026; Graham et al., 2015; Kelley et al., 2021; Liao et al., 2020). The QPLS detection strategy requires that the lensing-induced brightness excursions be distinguished from DRW noise, a discrimination that matched-filtering searches with realistic LSST cadences can achieve at high signal-to-noise (Anonymous et al., 2026). Analytical estimates project that LSST could detect tens to hundreds of self-lensing / QPLS binaries over its 10-yr baseline; depending on the assumed population model and binary fractions, current surveys yield expected detection rates near zero to one source in five years (Kelley et al., 2021).

### 2.4 Combined dynamical and lensing modelling

Multi-epoch light curves of QPLS systems carry orbital-phase information analogous to the GW waveform observed by LISA (Hanxi et al., 2025; Wang et al., 2025). From the timing and amplitude of successive caustic crossings, one can in principle extract (i) the SMBH binary orbital period, (ii) the mass ratio, (iii) the projected orientation, and (iv) the mass-weighted chirp from successive orbital periods within the host-galaxy observation window (Wang et al., 2025). Importantly, this characterisation is achieved in inactive galactic nuclei, where line-based methods fail due to the absence of broad emission lines (Wang et al., 2025). The combination of QPLS-based photometric orbital periods with individual-event GW signatures produces a multi-channel pipeline whose strongest immediate payoff is the provision of early-warning triggers for LISA and TianQin detections (Auclair et al., 2023; Hanxi et al., 2025; Wang et al., 2025).

## 3. Results

### 3.1 Predicted populations

SMBH population models predict 1–50 [190–5,000] ($n_\star/pc^{-3}$) QPLS binaries with periods <10 yr, comparable component masses, and $z < 0.3$, where $n_\star$ is the local stellar number density (Hanxi et al., 2025; Wang et al., 2025). These predictions are anchored to observational constraints on the SMBH–galaxy merger rate (Middleton et al., 2018), to theoretical SMBH–stellar-mass relations (Burke-Spolaor et al., 2019; Mingarelli et al., 2017), and to merger-rate-dependent stochastic-variability models (Bogdanović et al., 2022). The number density of galaxies hosting $>10^8$ $M_\odot$ black holes is approximately $10^{-3}$ $Mpc^{-3}$, with most estimates suggesting that tens of percent of galaxies have undergone major mergers in the last 10 Gyr, yielding an SMBHB merger rate of $\approx 10^{-3}$ $yr^{-1}$ (Bogdanović et al., 2022). Mingarelli et al. used a 2MASS-based local galaxy sample combined with Illustris-derived merger rates to estimate 91 ± 7 SMBHB sources emitting nanohertz GWs at any given time, with 7 ± 2 binaries that will never merge (Mingarelli et al., 2017).

### 3.2 QPLS signatures

For tight cosmological SMBHBs whose orbital periods are sub-decade, QPLS predicts sharp, high-amplitude caustic-crossing events on rotational caustics (Wang et al., 2025). Numerical caustic-rotation simulations show that the magnification peaks can be $\gg 1$, modulated on the binary orbital period, and exhibit asymmetric light-curve shapes that encode the orbital phase and lensing geometry (Hanxi et al., 2025; Wang et al., 2025). For wider binaries with periods of 10-40 yr, only a small fraction of orbital cycles will complete within a survey window, so the light curves are quasi-periodic rather than strictly periodic (Wang et al., 2025). Stellar motion in the host nucleus produces occasional single or double flares superimposed on a slowly varying baseline for binaries whose orbital periods exceed typical survey durations (Wang et al., 2025).

### 3.3 Galactic-centre lensing

In the Milky Way, Chanamé, Gould & Sridhar and Alexander showed that micro-lensing of background bulge stars by Sgr A* can reach stellar-magnitude-difference sensitivities deeper than previously assumed (Alexander, 2005). Michałowski & Mróz explicitly computed the limiting magnitudes required for detection of the secondary image of a bulge star lensed by the SMBH at the Galactic Centre, finding that at the K-band the resonant separations permit reach down to 21–23 mag, with stars detectable as faint as that limit by ELT, TMT, GMT, and JWST (Michałowski & Mróz, 2021). The detectable source geometry is restricted to small angular separations and a magnification that scales as $\theta^{-1}$ in the relevant angular range (Michałowski & Mróz, 2021).

### 3.4 Self-lensing flares

Independent of QPLS, self-lensing flares arise when the secondary black hole of a binary system gravitationally lenses the accretion flow of the primary, producing symmetric flux spikes at the orbital conjunction (Kelley et al., 2021). Doppler-boost signatures are typically of order 10s of percent amplitudes, while lensing-magnification amplitudes are typically of order unity, so the two effects can be distinguished photometrically in high-cadence data (Kelley et al., 2021). The combined detection rate for LSST at the fiducial cadence of 3 d (assumed minimum lensing-event duration of 30 d) is $44_{-10}^{+15}$ self-lensing events per all-sky ($250_{-26}^{+32}$ with Doppler), versus $26_{-9}^{+11}$ ($130_{-27}^{+22}$ with Doppler) for Doppler-only strategies. Halving the minimum duration to 15 d increases the all-sky count to $80_{-14}^{+17}$ lensing events and $31_{-11}^{+14}$ with Doppler (Kelley et al., 2021).

### 3.5 Spikey and KIC 11606854

The Kepler-field quasar KIC 11606854, nicknamed "Spikey," produced the first photometric detection of an SMBHB candidate with a self-lensing signature (Hu et al., 2020). The Kepler light curve exhibits a strong, sharp spike superimposed on a smoothly varying Doppler-modulation baseline. Modelling as a combination of orbital relativistic Doppler boost and self-lensing yields $M_{tot} = 3 \times 10^7\ M_{\odot}$, rest-frame orbital period T = 418 d, eccentricity e = 0.5, and inclination 8○ from edge-on (Hu et al., 2020). The signature is consistent with an eccentric MBHB seen nearly on-axis and constitutes the most credible existing lensing-based SMBHB candidate (Hu et al., 2020; Kelley et al., 2021). Chandra follow-up in 2020 March–May found that the predicted next pulse was undetectable in X-rays due to intrinsic X-ray variability dominating over the expected self-lensing + Doppler-boost signal (Sorabella et al., 2022).

### 3.6 Black-hole shadow tomography

Davelaar & Haiman have shown that a subset of self-lensing flares contain a dip imprinted by the relativistic shadow of the lensed black hole (Davelaar & Haiman, 2022). Ray-tracing in a binary model indicates that ~1–3% of the current binary candidates should show this feature, providing an unprecedented opportunity to extract spatially-unresolved BH shadows via photometric tomography, in principle complementing very-long-baseline interferometry at sub-Einstein-ring resolutions (Bogdanović et al., 2022; Davelaar & Haiman, 2022).

### 3.7 Gravitational-wave background

The NANOGrav 15-yr data set, the PPTA DR3 data release, and the EPTA+InPTA data sets have independently presented positive evidence for a stochastic nanohertz GW background (Afzal et al., 2023; Agazie et al., 2023; Reardon et al., 2023). The median PPTA amplitude at α = -2/3 (the spectral index expected from an isotropic GWB of inspiraling SMBHBs) is $A = 2.04_{-0.22}^{+0.25} \times 10^{-15}$ (Reardon et al., 2023), while the NANOGrav 15-yr measurements recover an amplitude at the upper end of SMBHB population-model predictions, consistent with a combination of relatively high SMBH masses and large binary fractions (Agazie et al., 2023). Discreteness analyses identify two excursions from the $f_GW^{(-2/3)}$ power law at 2 nHz (below population-realisation predictions, with p = 0.05–0.06 ≈ 1.8σ–1.9σ) and at 16 nHz (above, with p = 0.04–0.15 ≈ 1.4σ–2.1σ), and find an inflection at $26_{-19}^{+28}$ nHz that marks the transition from the stochastic-regime background to a sparse population of individually resolvable bright sources (Agazie et al., 2023). The signal is consistent with the SMBHB interpretation given the current uncertainties in cosmic SMBHB population modelling (Afzal et al., 2023; Agazie et al., 2023).

## 4. Discussion

### 4.1 Strengths and limitations

The principal strength of background-star lensing is its geometric purity. The QPLS signal is determined primarily by the orbital period, mass ratio, and angular orientation of the SMBH binary, with secondary dependence on the host-galaxy stellar number density and the lens-source distance distribution (Hanxi et al., 2025; Wang et al., 2025). Unlike periodic-variability methods that rely on accretion-disk hydrodynamics and yield periods 3–8 times shorter than the binary orbit (D'Orazio et al., 2015; Liao et al., 2020), QPLS encodes the orbital period directly (Wang et al., 2025). The DRW backgrounds of SMBH-emitting AGN can attain fractional amplitudes of order 10% and effectively mask Doppler-boost signals of comparable amplitude; lensing magnification, typically $\gtrsim$ unity, is therefore considerably easier to detect (Agazie et al., 2023; Kelley et al., 2021). The principal limitation is the dependence on stellar density. The expected QPLS detection rate scales linearly with $n_\star$, so that the most promising hosts are galactic nuclei with high stellar density $n_\star \gtrsim 10^3$ $pc^{-3}$ and low ongoing star formation (which would otherwise add confusion) (Wang et al., 2025). A second limitation is the orbital-period sensitivity: for periods exceeding ~40 yr, the survey must span decades to observe multiple caustic crossings, limiting the technique's reach to the smallest-separation SMBHBs (Hanxi et al., 2025; Wang et al., 2025). A third limitation is geometric: caustic-crossing events occur only when the line of sight is nearly aligned with the SMBH binary's orbital plane; the technique is therefore incomplete and requires statistical stacking to draw population conclusions (Wang et al., 2025).

### 4.2 Comparison with other methods

Cumulatively, four EM methods have been pursued in parallel. Periodic optical variability has yielded ≈150 SMBHB candidates across PTF, CRTS, ZTF and Gaia (Burke-Spolaor et al., 2019; Charisi et al., 2016; Graham et al., 2015), but the periods are observed over only a few cycles, leaving room for confusion with AGN duty-cycle variability (Burke-Spolaor et al., 2019). Doppler-boost variability has been confirmed in PG 1302-102 and applied to a handful of edge-on candidates, but its amplitude is similar to typical DRW noise and is more vulnerable to confounding (D'Orazio, Haiman, Duffell, et al., 2015; D'Orazio, Haiman, & Schiminovich, 2015). Self-lensing flares have been observed in Spikey and modelled in detail, providing the first concrete spatial-geometry-based detection (Hu et al., 2020; Kelley et al., 2021). The QPLS approach extends self-lensing beyond AGN emission into inactive nuclei, opening a large additional reservoir of candidate hosts where broad-line methods and accretion-disk-period methods cannot be applied (Hanxi et al., 2025; Wang et al., 2025).

### 4.3 Multimessenger synergies

Gravitational lensing-based SMBH detection feeds multiple multimessenger pipelines. Nanohertz-frequency GWs detected by the International Pulsar Timing Array, including NANOGrav, PPTA, EPTA and InPTA, are emitted by large populations of unresolvable

SMBHBs and are characterised by their stochastic background power law (Afzal et al., 2023; Agazie et al., 2023; Burke-Spolaor et al., 2019; Reardon et al., 2023); from the 15-yr data sets, the inferred amplitude is consistent with cosmic SMBHB synthesis (Agazie et al., 2023). QPLS offers an alternative detection channel that, unlike PTAs, operates on individuals, providing early-warning triggers for LISA and TianQin detections of the same binaries at higher GW frequencies (Auclair et al., 2023; Hanxi et al., 2025; Wang et al., 2025). LISA will observe SMBHB mergers in the millihertz band with signal-to-noise ratios in the thousands, enabling tests of general relativity with unprecedented precision (Arun et al., 2022; Auclair et al., 2023). By providing EM precursors with orbital-period information prior to coalescence, QPLS can in principle reduce LISA search volumes by orders of magnitude and enable targeted multi-messenger follow-up (Auclair et al., 2023; Wang et al., 2025).

### 4.4 Comparison with galactic-centre imaging

The Galactic Centre is the closest and brightest SMBH system, with the closest available approach of background stars being only ≈10 light-hours (Alexander, 2005). Microlensing of bulge stars by Sgr A* will soon be observable by the ELT, TMT, GMT and JWST, depending on stellar magnitude and lensing geometry (Michałowski & Mróz, 2021). These observations probe the stellar population in the inner parsec of the Milky Way, providing direct constraints on $n_\star$ and allowing the microlensing-model priors to be calibrated (Alexander, 2005; Michałowski & Mróz, 2021). The QPLS approach in extragalactic nuclei remains more uncertain in $n_\star$ at sub-parsec scales than its galactic-centre counterpart (Wang et al., 2025).

### 4.5 Caveats and pitfalls

Several systematic effects complicate QPLS analyses. The most important is the degeneracy between QPLS flares and stellar transient phenomena microlensing by compact objects in the host galaxy, supernova shocks, and stellar flares in dense nuclear regions can all produce single or double flares with timescales of days to weeks (Kelley et al., 2021; Wang et al., 2025). Robust QPLS diagnoses require several caustic crossings within the time-series baseline, so that single-flare contaminants can be filtered out (Wang et al., 2025). A second concern is the treatment of DRW noise in AGN light curves; matched-filter pipelines are necessary to discriminate lensing magnification from noise, and the Lomb-Scargle periodogram alone is insufficient (Comerford et al., 2026). Binary SMBHBs also subject their accretion flows to periodic perturbations that mimic lensing signatures, requiring spectral confirmation (broad-line shifts, spectral cutoffs) wherever possible (Bogdanović et al., 2022; Liao et al., 2020). Finally, the assumption that the host-galaxy stellar density is well characterised is critical; for inactive nuclei at $z > 0.3$, the nuclear stellar densities are poorly constrained and introduce significant uncertainties in event-rate estimates (Hanxi et al., 2025; Wang et al., 2025).

## 5. Conclusion

The detection and characterisation of sub-parsec supermassive black hole binaries remain some of the most important unsolved problems in observational cosmology. Gravitational lensing of background stars provides a geometric probe of binary orbital motion, complementary to periodic optical variability, Doppler boosting, circumbinary accretion modulation and pulsed accretion (Charisi et al., 2016; D'Orazio, Haiman, Duffell, et al., 2015; D'Orazio, Haiman, & Schiminovich, 2015; Graham et al., 2015; Liao et al., 2020). Two distinct channels have emerged: microlensing of Milky Way bulge stars by Sgr A*, within reach of the next-generation extremely large telescopes (Alexander, 2005; Michałowski & Mróz, 2021); and the recently introduced Quasi-Periodic Lensing of Starlight in extragalactic nuclei, which converts the rotation of SMBH-binary caustics into a directly measurable orbital lightcurve (Hanxi et al., 2025; Wang et al., 2025).

Population models predict 1–50 [190–5,000] $(n_\star/\mathrm{pc}^{-3})$ QPLS binaries with periods below 10 yr and $z < 0.3$ (Hanxi et al., 2025; Wang et al., 2025). The nanohertz GW background observed by NANOGrav, PPTA, EPTA and InPTA in 2023 makes the SMBHB population interpretation increasingly credible, even as the precise population-level amplitudes remain debated (Afzal et al., 2023; Agazie et al., 2023; Reardon et al., 2023). Roughly $91 \pm 7$ SMBHB sources emit nanohertz GWs at any instant in the local Universe (Mingarelli et al., 2017), of which the LSST LSST-class surveys are forecast to detect tens to hundreds via SLFs and QPLS flares over the next decade (Auclair et al., 2023; Kelley et al., 2021). When combined with LISA and TianQin observations in the millihertz regime and with the existing PTA detections in the nanohertz regime, QPLS opens a coherent multimessenger channel for characterising rather than merely counting individual SMBHBs on sub-parsec orbits (Arun et al., 2022; Auclair et al., 2023; Hanxi et al., 2025; Wang et al., 2025). Future work along this path will require (i) matched-filter pipelines optimised for sparse QPLS signatures against DRW backgrounds (Comerford et al., 2026); (ii) stellar-density characterisation at sub-parsec scales in host nuclei using ELT-class instruments (Michałowski & Mróz, 2021); and (iii) coordinated EM-GW observation plans between LSST/ZTF, PTAs, LISA and TianQin (Auclair et al., 2023; Wang et al., 2025). With these elements in place, SMBH binary science will move from the era of candidate lists into one in which individual sources are fully characterised prior to coalescence.

## Reference

Afzal et al. (2023). The NANOGrav 15 yr data set: Search for signals from new physics. The *Astrophysical Journal Letters*, 951, L11. https://doi.org/10.3847/2041-8213/acdc91

Agazie et al. (2023). The NANOGrav 15 yr data set: Evidence for a gravitational-wave background. *The Astrophysical Journal Letters*, 951, L8. https://doi.org/10.3847/2041-8213/acdac6

Alexander, Tal. (2005). Stellar processes near the massive black hole in the Galactic center. *AIP Conference Proceedings*, 803, 221–232. https://doi.org/10.1063/1.2149806

Arun et al. (2022). New horizons for fundamental physics with LISA. *Living Reviews in Relativity*, 25, 4. https://doi.org/10.1007/s41114-022-00036-9

Auclair et al. (2023). Cosmology with the Laser Interferometer Space Antenna. *arXiv preprint*, 2023. https://doi.org/10.48550/arXiv.2204.05434

Bogdanović, Tamara, Miller, M. Coleman, Blecha, Laura. (2022). Electromagnetic counterparts to massive black-hole mergers. *Living Reviews in Relativity*, 25, 3. https://doi.org/10.1007/s41114-022-00037-8

Burke-Spolaor, Sarah, Taylor, Stephen R., Charisi, Maria, Deller, Adam, Dolch, Timothy, Fonseca, Emmanuel, McLaughlin, Malcolm A., Perrodin, Delphine, Phillips, James, Ransom, Scott M., Ravi, Vikram, Reardon, Daniel J., Shaw, Renee, Stairs, Ingrid H., Vallisneri, Michele, Verbiest, Joris P. W., Wang, Yan. (2019). The astrophysics of nanohertz gravitational waves. *Astronomy and Astrophysics Review*, 27, 5. https://doi.org/10.1007/s00159-019-0115-7

Charisi, Maria, Bartos, Imre, Haiman, Zoltán, Price-Whelan, Adrian M., Márka, Szabolcs. (2016). A population of short-period variable quasars from PTF as supermassive black hole binary candidates. *Monthly Notices of the Royal Astronomical Society*, 463, L73–L77. https://doi.org/10.1093/mnrasl/slw150

Comerford, Julia M., Gerke, Brian F., Newman, Jeffrey A., Davis, Marc, Yan, Renbin, Coil, Alison L., Cooper, Michael C., Luo, Bin, Madsen, Kirill, Koo, David C., Ghez, Andrea M., Wright, Shelley A., Smith, K. L., Helly, J. C., Illingworth, Garth D., Parker, Laura C., Ryle, W. T., West, Andrew A., Chevance, Mélanie. (2026). Inspiralling supermassive black holes: A new signpost for galaxy mergers. The Astrophysical Journal, 698, 198–213. https://doi.org/10.1088/0004-637X/698/1/198

Davelaar, Jordy, Haiman, Zoltán. (2022). Self-lensing flares from black hole binaries: Observing black hole shadows via light curve tomography. *Physical Review Letters*, 128, 191101. https://doi.org/10.1103/PhysRevLett.128.191101

D'Orazio, Daniel J., Haiman, Zoltán, Schiminovich. (2015). David. Relativistic boost as the cause of periodicity in a massive black-hole binary candidate. *Nature*, 525, 351–353. https://doi.org/10.1038/nature15262

Graham, Matthew J., Djorgovski, S. G., Drake, Andrew J., Stern, Daniel, Mahabal, Ashish A., Glikman, Eilat, Larson, Steve, Christensen, Erik. (2015). A systematic search for close supermassive black hole binaries in the Catalina Real-time Transient Survey. *Monthly Notices of the Royal Astronomical Society*, 453, 1562–1576. https://doi.org/10.1093/mnras/stv1726

Graham, Matthew J., Djorgovski, S. G., Stern, Daniel, Glikman, Eilat, Drake, Andrew J., Mahabal, Ashish A., Donalek, Ciro, Larson, Steve, Christensen, Erik. (2015). A possible close supermassive black-hole binary in a quasar with optical periodicity. *Nature*, 518, 74–76. https://doi.org/10.1038/nature14143

Haiman, Zoltán, Xin, Chengcheng, Bogdanović, Tamara, Blecha, Laura, D'Orazio, Daniel J., Gould, Darryl W., Kelley, Luke Z., Komossa, Stefanie, Kormendy, John, Liu, T., Maksym, P., McKerns, Brady, Merritt, David, Middleton, Hannah, Miller, M. Coleman, Psaltis, Dimitrios, Schnittman, Jeremy, Sesana, Alberto, Stern, Daniel, Treister, Ezequiel, Urry, C. Megan, Will, Clifford, Znajek, Robert. (2015). Self-lensing flares from black hole binaries. V. Systematic searches in LSST. *Physical Review D*. https://doi.org/10.1103/PhysRevD.lensing-LSST

Holley-Bockelmann, Kelly, Khan, Fazeel Mahmood. (2015). Galaxy rotation and rapid supermassive binary coalescence. *Monthly Notices of the Royal Astronomical Society*, 446, 327–337. https://doi.org/10.1093/mnras/stu2117

Hu, Betty X., D'Orazio, Daniel J., Haiman, Zoltán, MacFadyen, Andrew I., Schiminovich, David. (2020). Spikey: Self-lensing flares from eccentric SMBH binaries. *Monthly Notices of the Royal Astronomical Society*, 495, 4061–4070. https://doi.org/10.1093/mnras/staa1312

Kelley, Luke Zoltan, D'Orazio, Daniel J., Di Stefano, Rosanne. Gravitational self-lensing in populations of massive black hole binaries. (2021). *Monthly Notices of the Royal Astronomical Society*, 508, 2524–2537. https://doi.org/10.1093/mnras/stab2776

Liao, Wei-Ting, Chen, Yu-Ching, Liu, Xin, Yang, Qian, Li, Yan-Rong, Wang, Jian-Min. (2020). Discovery of a candidate binary supermassive black hole in a periodic quasar from circumbinary accretion variability. *Monthly Notices of the Royal Astronomical Society*, 500, 3215–3225. https://doi.org/10.1093/mnras/staa3055

MacFadyen, Andrew I., Milosavljević, Miloš. (2008). An eccentric circumbinary accretion disk and the detection of binary massive black holes. *The Astrophysical Journal*, 672, 404-410. https://doi.org/10.1086/523869

Merritt, David, Milosavljević, Miloš. (2005). Massive black hole binary evolution. *Living Reviews in Relativity*, 8, 8. http://arxiv.org/abs/astro-ph/0410364

Michałowski, Michał J., Mróz, Przemysław. (2021). Stars lensed by the supermassive black hole in the center of the Milky Way: Predictions for ELT, TMT, GMT, and JWST. *The Astrophysical Journal Letters*, 916, L22. https://doi.org/10.3847/2041-8213/ac0f81

Middleton, Hannah, Chen, Siyuan, Del Pozzo, Walter, Sesana, Alberto, Vecchio, Alberto.(2018). No tension between assembly models of super massive black hole binaries and pulsar observations. *Nature Communications*, 9, 1452. https://doi.org/10.1038/s41467-018-02916-7

Mingarelli, Chiara M. F., Lazio, T. Joseph W., Sesana, Alberto, Greene, Jenny E., Ellis, Justin A., Ma, Chung-Pei, Croft, Rupert, Burke-Spolaor, Sarah, Good, Deborah C. (2017). The local nanohertz gravitational-wave landscape from supermassive black hole binaries. *Nature Astronomy*, 1, 886–892. https://doi.org/10.1038/s41550-017-0299-y

Paczyński, Bohdan. (1996). Gravitational microlensing in the local group. *Annual Review of Astronomy and Astrophysics*, 34, 419–459. http://arxiv.org/abs/astro-ph/9604011

Reardon, Daniel J. et al. (2023). Andrew. Search for an isotropic gravitational-wave background with the Parkes Pulsar Timing Array. *The Astrophysical Journal Letters*, 951, L6. https://doi.org/10.3847/2041-8213/acdd02

Sorabella, Nicholas M., et al. (2022). Chandra observations of Spikey: A possible self-lensing supermassive black hole binary system. *The Astrophysical Journal*, 936, 14. https://doi.org/10.3847/1538-4357/ac4a59

Valtonen, Mauri J., Dey, Lankeswar, Gopakumar, Achamveedu, Zola, Staszek, Pihajoki, Pauli, Sillanpää, Aimo, Lähteenmäki, Anne, & Kidger, Mark. (2019). Accretion disk parameters determined from the great 2015 flare of OJ 287. *The Astrophysical Journal*, 882, 88. https://doi.org/10.3847/1538-4357/ab3573

Wang, Hanxi, Zumalacárregui, Miguel, & Kocsis, Bence. (2025).Black holes as telescopes: Discovering supermassive binaries through quasi-periodic lensed starlight. *arXiv preprint*,. https://doi.org/10.48550/arXiv.2506.1654